\documentclass[noshowpacs,amsmath,twocolumn,
aps,prb]
{revtex4-1}

\usepackage{setspace}
\usepackage{amsmath}
\usepackage{graphicx}
\usepackage[nearskip,margin = 0pt]{subfig}
\usepackage{subfig}

\usepackage{verbatim}
\usepackage{amsfonts}
\usepackage{amssymb}
\usepackage{epstopdf} 
\usepackage{xcolor}
\DeclareGraphicsExtensions{.pdf,.eps,.png,.jpg,.mps} 

\begin{document}

\title{Soliton Microcomb Range Measurement}

\author{Myoung-Gyun Suh and Kerry Vahala$^{*}$\\
T. J. Watson Laboratory of Applied Physics, California Institute of Technology, Pasadena, California 91125, USA.\\
$^*$Corresponding author: vahala@caltech.edu
}

\maketitle
\newcommand{\ts}{\textsuperscript}
\newcommand{\tsb}{\textsubscript}



{\bf
Laser-based range measurement systems (LIDAR) are important in many application areas including autonomous vehicles, robotics, manufacturing, formation-flying of satellites, and basic science. Coherent laser ranging systems using dual frequency combs provide an unprecedented combination of long range, high precision and fast update rate. Here, dual-comb distance measurement using chip-based soliton microcombs is demonstrated. Moreover, the dual frequency combs are generated within a single microresonator as counter-propating solitons using a single pump laser. Time-of-flight measurement with 200 nm precision at 500 ms averaging time is demonstrated. Also, the dual comb method extends the ambiguity distance to 26 km despite a soliton spatial period of only 16 mm. This chip-based source is an important step towards miniature dual-comb laser ranging systems that are suitable for photonic integration. 
}



The invention of the optical frequency comb has had a major impact on laser ranging systems. In addition to providing a highly accurate frequency calibration source in methods such as multi-wavelength interferometry\cite{salvade2008high} and frequency-modulated continuous wave laser interferometry\cite{baumann2013comb}, frequency combs have enabled a new method of ranging called dual-comb LIDAR\cite{coddington2009rapid} (DCL).  In this method, two frequency combs having slightly different repetition rates are phase locked. In the detection process, the resulting intercomb beats create a repetitive interferogram that is able to attain sub-nanometer range precision. At the same time, the ambiguity range of the combined comb system is greatly extended beyond the pulse-to-pulse separation distance of each comb. Applications of DCL systems would benefit from more compact and miniature comb systems, and the recent development of a miniature frequency comb (microcomb \cite{del2007optical,kippenberg2011microresonator}) suggests that chip-integrated DCL systems may be possible. Microcombs have been demonstrated in several material systems\cite{del2007optical,grudinin2009generation,papp2011spectral,okawachi2011octave,li2012low,hausmann2014diamond} and many implementations are monolithic on a silicon wafer so that integration with both other photonic components\cite{razzari2010cmos,ferdous2011spectral,jung2013optical} as well as electronics is possible. If applied to DCLs, full integration would enable scalable manufacturing for mass market applications. 

A recent advancement in microcombs has been the realization of soliton mode-locking\cite{herr2014temporal,yi2015soliton,brasch2016photonic,wang2016intracavity,joshi2016thermally}, which provides phase-locked femtosecond pulses with GigaHertz to TeraHertz repetition rates. Soliton microcombs are being studied in several frequency comb applications, including optical frequency synthesis\cite{spencer2017towards}, secondary time standards \cite{Frank2017clock}, and dual comb spectroscopy\cite{suh2016microresonator,dutt2016chip,pavlov2017soliton}. In this work, we demonstrate time-of-flight distance measurement using a chip-based dual-soliton source. Beyond the demonstration of microcomb LIDAR, the two soliton streams are generated as counter-propagating solitons within a single resonator\cite{yang2017counter}. This simplifies the system by eliminating the need for two resonators and two pump lasers. It also improves the mutual coherence between the two combs.

\begin{figure*}
    \begin{centering}
   \includegraphics[width=\linewidth]{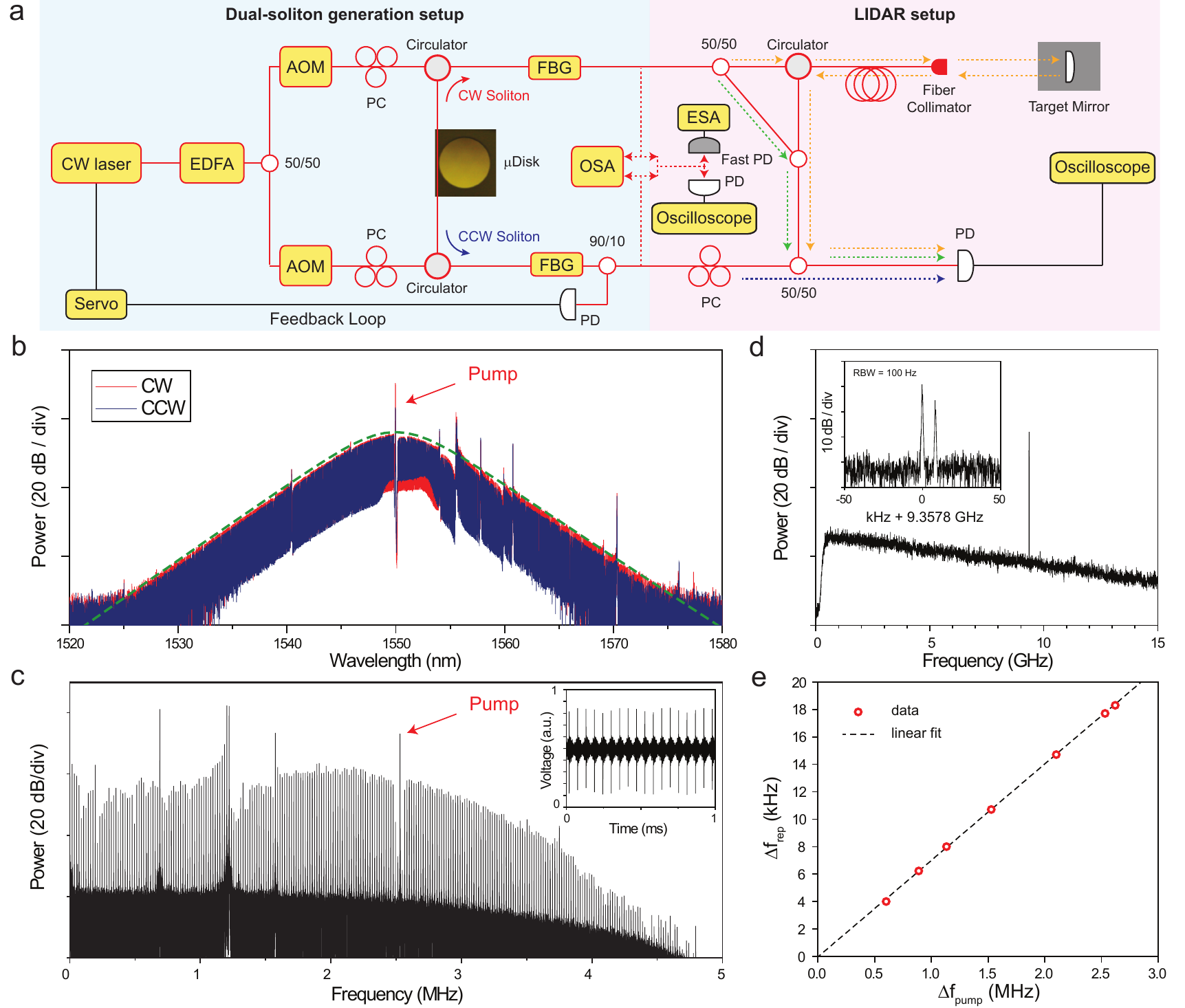}       \captionsetup{singlelinecheck=no, justification = RaggedRight}
        \caption{{\bf Experimental setup for dual-soliton generation and ranging measurement}. (a) The experimental setup consists of a dual-soliton generation setup (blue panel) and ranging setup (red panel). A fiber laser at 1550 nm is amplified by an erbium-doped fiber amplifier (EDFA) and split using a 50/50 coupler to pump the resonator in two directions. In each path, an acousto-optic modulator (AOM) is used to modulate the pump power for soliton triggering and to tune the pump frequency. The pump light is coupled into the microresonator by evanescent coupling via a fiber taper, and the polarization of the pump laser is adjusted using a polarization controller (PC). Solitons are triggered in both CW and CCW direction, and stabilized by a servo feedback loop using the soliton power detected at a photodetector (PD) as an error signal\cite{yi2016active}. A fiber bragg grating (FBG) filter is used to attenuate the transmitted pump power. An optical spectrum analyzer (OSA) and electrical spectrum analyzer (ESA) are used to analyze the dual-soliton source. For distance detection, the CW soliton stream is split into two paths: the reference path (green dotted line) and the target path (orange dotted line). The target path includes a circulator, fiber delay line, fiber collimator and mirror. The CW soliton streams of both paths are combined with the CCW soliton stream (blue dotted line) and photodetected. (b) Typical optical spectra of the CW/CCW solitons. Hyperbolic-secant-square fit (green dotted curve) using a soliton pulse width of 200 fs as a fitting parameter is overlaid onto the spectra. (c) Electrical spectrum of heterodyned dual-soliton source with $\Delta f_{rep}$ $\sim$ 18 kHz. Spectrum is obtained by Fourier-transforming the photodetected interferogram (inset). Interferogram is measured for 100 ms and a 1 ms time span is shown. Resolution bandwidth (RBW) is 10 Hz. Heterodyne frequency component of the two pump lasers is indicated near 2.53 MHz. (d) Electrical spectrum of the solitons showing the approximate repetition frequency $f_{rep}$ $\sim$ 9.36 GHz. Inset is a zoom-in of the spectrum showing the resolved CW/CCW repetition rates. (e) Repetition frequency difference ($\Delta f_{rep}$) versus pump frequency difference ($\Delta f_{pump}$). }
    \end{centering}
\end{figure*}


\begin{figure*}[t!]
    \begin{centering}
  \includegraphics[width=\linewidth]{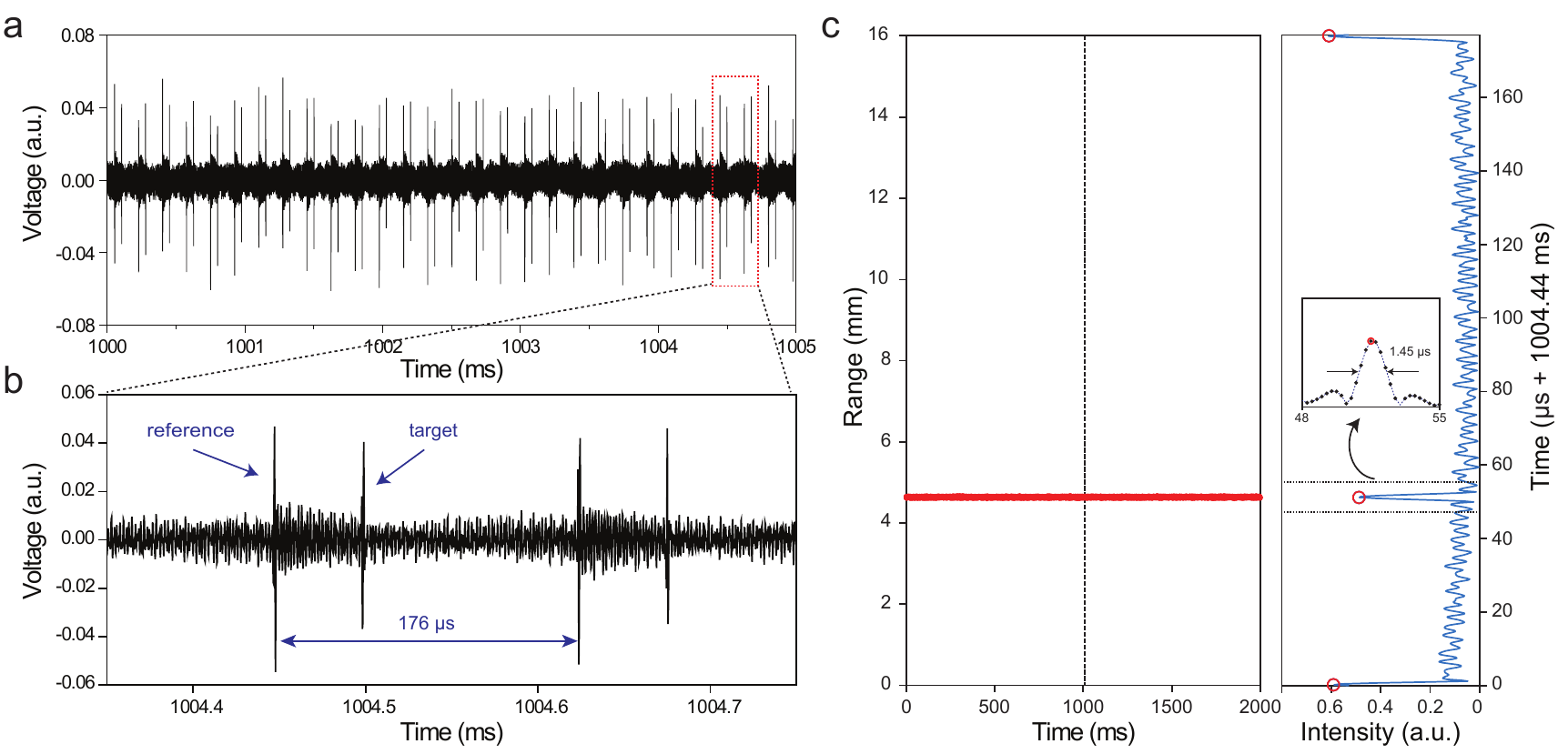}
        \captionsetup{singlelinecheck=no, justification = RaggedRight}
        \caption{{\bf Distance measurement} (a) Typical interferogram containing range information. (b) Zoom-in of the interferogram over two time periods. The reference peaks and target peaks are shown and the time period is approximately 176 $\mu$s.  (c) The measured distance between the reference peak and target peak is plotted versus time in the left panel. The range ambiguity in this measurement is 16 mm. Right panel shows the electrical intensity trace (blue) and peaks (red circle) for a data trace near 1 s (dashed vertical line in left panel). Inset: zoom-in of the target intensity peak showing the electrical pulse width of 1.45 $\mu$s. }
\end{centering}
\end{figure*}

The experimental setup is shown in figure 1a. As described there, two pump fields are coupled to the microresonator along the clockwise (CW) and counterclockwise (CCW) directions of a common whispering-gallery resonance. The microresonator is a silica wedge disk fabricated on a silicon wafer\cite{lee2012chemically}. The resonator had an unloaded quality factor of approximately 300 $\sim$ 500 million and a 7 mm diameter corresponding to a 9.36 GHz free spectral range (FSR). With the high circulating power, frequency combs are initiated by way of parametric oscillation \cite{kippenberg2004kerr,savchenkov2004low} and are broadened by cascaded four-wave mixing\cite{del2007optical, kippenberg2011microresonator}. Solitons are generated in both CW and CCW directions and stabilized using a feedback loop as described elsewhere\cite{yi2016active}. The servo control holds the frequency detuning of one pump direction fixed relative to the cavity resonant frequency, while the second pump frequency can be independently tuned using an AOM. The generated CW/CCW soliton streams are coupled in opposing directions to the optical fiber and then transferred towards the LIDAR setup by using circulators. 

To characterize the solitons the CW and CCW solitons are also combined by tapping power from the main fiber using the dotted paths in figure 1a. Superimposed optical spectra of the CW and CCW solitons (measured on the OSA) are shown in figure 1b. The characteristic hyperbolic-secant-square function (green dotted curve) is fit to the spectral envelope and a soliton pulse width of 200 fs is determined from this fitting. The combined soliton streams were also detected by two PDs. The output voltage of one of the PDs measured using an oscilloscope is shown in the figure 1c inset. The periodic pulses in this time trace reflect the periodic interference of the dual soliton pulse streams as they stroboscopically interfere on account of the slight difference in their respective repetition rates ($\Delta f_{rep}$ $\sim$ 18 kHz).  This voltage time trace signal (or interferogram) is Fourier transformed to obtain the electrical spectrum in figure 1c main panel. The CW and CCW pumps were offset in frequency by $\Delta f_{pump}$ = 2.53 MHz in this measurement and their corresponding beatnote is indicated in figure 1c main panel. The linewidth of the pump beat note is approximately 20 Hz and is limited by noise in the AOM driver used to create the two pump frequencies (see figure 1a). The linewidths of the interferogram spectral lines are observed to become continuously narrower as the frequency of spectral lines decreases from $\Delta f_{pump}$ towards zero frequency. This decreasing linewidth results from the phase locking of the CW and CCW solitons and their relative repetition rates\cite{yang2017counter}.

Using a fast PD (50 GHz bandwidth), the electrical power spectrum of the detected dual soliton pulse streams was measured on the ESA (figure 1d).  The repetition frequency ($f_{rep}$) is approximately 9.36 GHz. When the electrical spectrum is zoomed-in near $f_{rep}$ (fig. 1d inset), the two distinct repetition frequencies of the CW/CCW solitons are resolved. The repetition frequency difference ($\Delta f_{rep}$) between CW and CCW solitons is adjusted by tuning the frequency difference of the two pump lasers ($\Delta f_{pump}$) using the AOMs. This tuning has been shown to result from the Raman-induced soliton self-frequency shift\cite{yang2017counter}, and typically, $\Delta f_{rep}$ increases with increasing $\Delta f_{pump}$. The maximum $\Delta f_{rep}$ $\sim$ 20 kHz is achieved when $\Delta f_{pump}$ $\sim$ 3 MHz as shown in figure 1e. Here, the maximum $\Delta f_{pump}$ is limited by the 3dB frequency shift range of the AOM. 


The two soliton streams are coupled into the LIDAR setup shown in figure 1a (right red panel). In the figure, the CW soliton stream is used to acquire the distance information. It is split into a reference path (green dotted arrow) and target path (orange dotted arrow) via the 50/50 coupler.  The target path (CW) soliton stream also passes through a circulator on its way to a gradient-index (GRIN) collimator. The collimator emits the soliton pulses towards a target mirror. The CW soliton stream reflected from the target is collected by the collimator and then combined with the reference CW soliton stream. A fiber delay line (approximately 15 m physical path length) was added before the collimator to increase the effective target distance. The distance difference between the reference path and the target path includes a roundtrip through the fiber delay and the free-space path. It also includes the two fiber segments at the circulator (see figure 1a). In the data, this distance difference is divided by two. Finally, the CW soliton stream carrying the distance information (green/orange dotted arrow) is combined with CCW soliton stream (blue dotted arrow) to generate the interferogram. The polarization of the CCW soliton stream is adjusted to maximize the target reflection peaks in the interferogram.

\begin{figure}[t!]
  \begin{centering}
  \includegraphics[width=\linewidth]{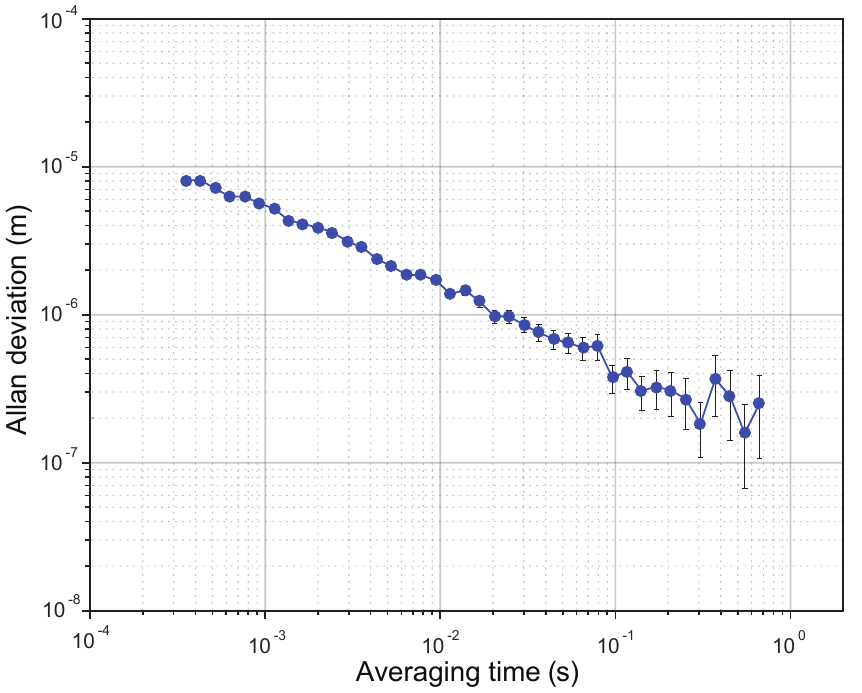}
    \captionsetup{singlelinecheck=no, justification = RaggedRight}
    \caption{{\bf Precision of distance measurement versus averaging time.} Allan deviation is calculated from the 2 s time series of measured distance (i.e., reference peak to signal peak). 200 nm precision is achieved near 500 ms averaging time. Error range is calculated from the Allan deviation divided by square root of sample size.}
    \label{fig3}
  \end{centering}
\end{figure}

Figure 2a shows a portion of the measured LIDAR interferogram as displayed by the oscilloscope. The interferogram is recorded for a total of 2 seconds with a 250 ns sampling time. Only a 5 ms portion of the overall span is shown in the figure. Figure 2b is a zoomed-in view of the interferogram showing the reference peaks and target peaks within two periods of 176 $\mu$s, corresponding to $\Delta f_{rep}$ $\sim$ 5.685 kHz. The temporal location of the peaks in the electrical pulse stream is determined by Hilbert transforming the interferogram. The time interval between a reference peak and a target peak is then calculated for each period. The time interval at each period is converted to the distance scale and plotted in figure 2c. The time increment in the plot is the interferogram period (176 $\mu$s). The averaged target distance is 4.637429 mm. The ambiguity range is 16 mm, corresponding to 1/2 of the pulse-to-pulse distance separation at the 9.36 GHz soliton repetition rate. The right panel of figure 2c shows one time period of the Hilbert transformed interferogram near 1 s in the measurement (the vertical dotted line in left panel). The inset of the right panel gives the zoomed-in target intensity peak and shows the full-width half-maximum pulse width of $\sim$ 1.45 $\mu$s.

Figure 3 studies the precision of the time-of-flight distance measurement. The Allan deviation of the time-series distance is calculated at averaging times ranging from 352 $\mu$s to 667 ms. By fitting the plot in figure 3, the measured precision is $ \sigma$ $\sim$ 10 $\mu$m $(T_{update}/T)^{1/2}$, where $T_{update}$ $\sim$ 176 $\mu$s is the update time and $T$ is the averaging time. Near 500 ms averaging time, a precision of 200 nm is achieved.

\begin{figure}[t!]
  \begin{centering}
  \includegraphics[width=\linewidth]{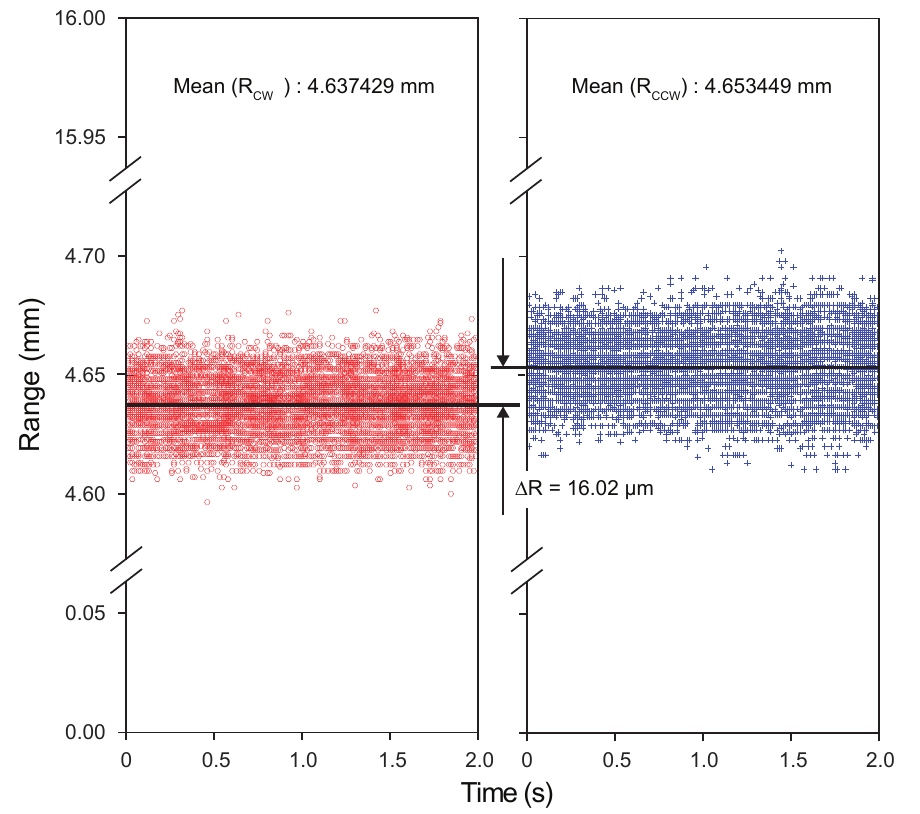}
  \captionsetup{singlelinecheck=no, justification = RaggedRight}
    \caption{{\bf Resolving range ambiguity.} $R_{CW}$ (red data) and $R_{CCW}$ (blue data) are range versus time using the CW and CCW solitons, respectively. The distance difference ($\Delta R$ = Mean($R_{CW}$) - Mean($R_{CCW}$)) between the two measurements is 16.02 $\mu$m and is used to determine the absolute range (see text).}
  \end{centering}
\end{figure}

The distance measurement shown in figure 2c has a range ambiguity of 16 mm. To greatly extend this ambiguity range, a similar distance measurement was performed after swapping the roles of two soliton streams.  As described elsewhere, the Vernier effect resulting from the difference in the soliton repetition rates can then be used to resolve the range ambiguity\cite{coddington2009rapid}. Figure 4 shows a zoom-in view of the two distance measurements where $R_{CW}$ (red circle) and $R_{CCW}$ (blue cross) are the measured distances when the CW soliton and the CCW soliton are used for ranging. The average distance difference, $\Delta R$ = Mean($R_{CW}$) - Mean($R_{CCW}$), between the two measurements is 16.02 $\mu$m. Considering the Vernier effect\cite{coddington2009rapid}, the ambiguity-resolved distance is $R' = \Delta R (f_{rep}^{CCW}/\Delta f_{rep}) + R_{CW}$ $\approx$ 26.3729 m $\pm$ 0.466 m with a new ambiguity range of $\sim$ 26 km. This measured distance is in good agreement with the optical path length of 26.815 m of the target including fiber delay as measured using an optical time-domain reflectometer (Luna OBR 4400). The uncertainty of ($\pm$ 0.466 m) in this ambiguity resolved measurement results from the original 200 nm precision multiplied by $\sqrt{2} f_{rep}/\Delta f_{rep}$. If this uncertainty were below the pulse-to-pulse range ambiguity, which would require a precision of $ (c \Delta f_{rep})/(2\sqrt{2} f_{rep}^2) $  (6.88 nm in this experiment), then a nanometer-scale precision would be possible in the ambiguity-resolved case\cite{coddington2009rapid}. Because the precision of the current measurement is limited by noise in the electrical driver of the AOMs, using a lower noise driver could dramatically improve the precision. Complementary interferometric measurement in addition to the time-of-flight could also improve the precision. Finally, increasing $\Delta f_{rep}$ makes it easier to achieve better precision in the ambiguity resolved case, and it allows greater averaging of distance data over a given interval which also improves precision.

In current work, the update time of $\sim$ 176 $\mu$s and the ambiguity range of $\sim$ 26 km are determined by $\Delta f_{rep}$ = 5.685 kHz, which can be tuned by changing $\Delta f_{pump}$. The tunability of $\Delta f_{rep}$ makes the dual soliton source interesting in that the LIDAR system can be adjusted to provide an optimal update time and ambiguity range according to the application. For example, in applications requiring faster update rate, increasing $\Delta f_{rep}$ to 1 MHz can improve the update time to 1 $\mu$s with a reduced ambiguity range of 150 m.

In summary, a soliton dual comb LIDAR system has been demonstrated using a single, chip-based microresonator pumped by a single laser. Sources like this can be be used to miniaturize and simplify conventional dual-comb ranging systems. Moreover, using a waveguide-integrated structure\cite{yang2017integrated} to generate the solitons, the chip-based dual-soliton source can be integrated with other on-chip optical components including elements necessary to create an optical phased array\cite{sun2013large} for beam steering. 

\vspace{3 mm}

Note: The authors would like to draw the readers' attention to other soliton microcomb range measurement work\cite{ganin2017ultrafast}, which was reported while preparing this manuscript.

\vspace{3 mm}


\noindent 

\vspace{3 mm}

\noindent \textbf{Acknowledgments}
The authors gratefully acknowledge the Defense Advanced Research Projects Agency under the PULSE program (Award No. W31P4Q-14-1-0001). The authors also thank the Kavli Nanoscience Institute.
\vspace{1 mm}

\noindent\textbf{Author Contributions} 
MGS and KV conceived the experiment. MGS fabricated devices and conducted the measurements. MGS and KV analyzed the data and wrote the manuscript.
\vspace{1 mm}

\noindent \textbf{Author Information} Correspondence and requests for materials should be addressed to KJV (vahala@caltech.edu ).

\bibliography{main}

\end{document}